# Reading the sky
# And
# The spiral of teaching and learning in astronomy

Urban Eriksson[†1,2,*]

[1]Kristianstad University, 291 88, Kristianstad, Sweden
[2]The National Resource Center for Physics Education, Lund University, Box 118, 221 00 Lund, Sweden

This theoretical paper introduces a new way to view and characterize teaching and learning astronomy. It describes a framework, based on results from empirical data, analyzed through standard qualitative research methodology, in which a theoretical model for vital competencies of learning astronomy is proposed: *Reading the Sky*. This model takes into account not only *disciplinary knowledge* but also *disciplinary discernment* and *extrapolating three-dimensionality*. Together, these constitute the foundation for the competency referred to as *Reading the Sky*. In this paper, I describe these concepts and how I see them being connected and intertwined to form a new competency model for learning astronomy and how this can be used to inform astronomy education to better match the challenges students face when entering the discipline of astronomy: *The Spiral of Teaching and Learning*. Two examples are presented to highlight how this model can be used in teaching situations.

*'You see, but you do not observe. The distinction is clear.'*
Sherlock Holmes, in 'A Scandal in Bohemia', Arthur Conan Doyle (1891)

*'A lily is more real to a naturalist than it is to an ordinary person. But it is still more real to a botanist. And yet another stage of reality is reached with that botanist who is a specialist in lilies.'*
Nabokov (1962)

## I. INTRODUCTION

Learning astronomy could be exciting but also challenging and demanding for many students. Over the years, many papers have been published describing various difficulties students face and present when learning astronomy [see, for excellent reviews, 1,2]. These difficulties often revolve around astronomical concepts that in astronomy courses are being presented using a multitude of different disciplinary-specific *semiotic resources*, including *representations, tools and activities* [3]. Also, learning astronomy involves being able to think about these astronomical concepts in three or four dimensions (3D, 4D). Astronomy as a discipline is special; it is not possible to directly access the Universe by one's eyes, except for the Moon, the stars in the night sky, etc. Instead, every bit of information about the Universe is gathered by different *tools*, i.e. telescopes and detectors, and processed, and finally presented using different types of semiotic resources. Learning astronomy then becomes learning to interpret these semiotic resources, which can be seen as learning a new language; it has its particular language and "grammar", i.e. a novice needs to learn how to read, write, and use all the different disciplinary-specific semiotic resources that constitute the disciplinary discourse of astronomy [3-5]. Moreover, research has shown that multidimensional (MD) thinking of space, or *extrapolating three-dimensionality* (E3D) from one- or two-dimensional semiotic resources, see below, is both very important and very difficult for students to master [6-10]. These difficulties taken into account makes it challenging for new-to-the-discipline students to learn astronomy, since not only does the student need to learn disciplinary knowledge, but (s)he needs to learn to "read" and "write" all the different highly specialised disciplinary-specific semiotic resources that astronomers use to communicate within the discipline [4,5,11-13]. I refer to this "reading" of disciplinary-specific semiotic resources as *disciplinary discernment* and below I describe how this discernment can be characterized by the Anatomy of Disciplinary Discernment (ADD) [14].

Below, I combine these educational ideas and findings to define and discuss a new competency important for teaching and learning of astronomy – *Reading the Sky* [10,15]. Using this as a starting point, I introduce a new framework for how to inform teaching and learning astronomy: *The Spiral of Teaching and Learning* (STL).

### A. Aim and Research question

Based on the above, the aim for this paper is to introduce, discuss and problematize:

---

[*] Corresponding author: urban.eriksson@hkr.se

➢ What is *Reading the Sky* competency and how can it be used to inform teaching and learning astronomy?

After this introduction, I go on to present the background and theoretical framework, which includes a discussion on competency, a summary of the ADD and E3D, which leads to the concept of *reading* in different context. In the following sections I describe *Reading the Sky* and how I see it as a competency. After this I introduce the *Spiral of Teaching and Learning*, followed by a discussion where I present implications from teaching, using two examples, before the conclusions.

## II. BACKGROUND AND THEORETICAL FRAMEWORK

Below I describe in detail the fundamental concepts which the framework *Reading the Sky* is built upon.

### A. Competency - a definition

Today, the value of knowledge lies much more in competence performativity and innovation than in simply knowing. As such, education today is increasingly moving towards a competence mindset. Reaching such competency in areas such as science has long been known to be challenging [see, for example, 16,17,18], but no one have addressed this from a learning astronomy perspective. Moving from everyday conceptions of the world around us to disciplinary interpretations are fraught with pitfalls and problems. Thus, it becomes an important educational consideration to understand what underpins the competency characteristics of the disciplinary trajectory.

Therefore, to move the augmentation forward, a definition of what is meant by *competency* is needed. For the purpose presented here, I draw on David Dubois' [19] definition of competency as:

> 'those characteristics–knowledge, skills, mindsets, thought patterns, and the like– that when used whether singularly or in various combinations, result in successful performance' (p. v).

Before building the concept of *Reading the Sky*, I will now introduce two essential concepts mentioned above and central to *Reading the Sky*. These are: *The Anatomy of Disciplinary Discernment* [14] and, *Extrapolating Three-Dimensionality* [6].

### B. The Anatomy Disciplinary Discernment (ADD)

At any time in our daily life, humans are exposed to huge amounts of information though our senses, but one can only focus on a small portion of this information at a time [20]. The big questions are what to focus on, and discern, and how do one know what is important? Here, I characterize this as *discernment* in terms of coming to know what to focus on and how to appropriately interpret it for a given context. Becoming competent in any discipline involves a similar process, namely learning [14]; '*what* to focus on in a given situation and *how* to interpret it in an appropriate, disciplinary manner' (p. 168). This involves two concepts, *noticing* and *reflection*, which is used to define disciplinary discernment. *Noticing* is connected to learning by experiencing new things or by new observations that trigger new ideas. In astronomy, this happens mostly through visual perception, by noticing of something from a disciplinary-specific representation [3]. The process of noticing is an unconscious act by humans we cannot control [21]. 'Our senses provide information to our brain that we process, usually in an unconscious way, and only some of this information comes to our conscious awareness' [14](p. 168), i.e. to distinguish it from the background. For humans to remember something, we need to mark it in our working memory, and can then use it for different things [21], by taking it back into focal awareness to construct meaning [22], hence change one's thinking. This meaning-making characterizes the process of learning by discernment.

Now, one would think that everyone can notice the same things from a representation. This is not the case since the noticing depends on one's earlier experiences, background and disciplinary educational level [23,24]. The connection to learning is referred to by Lindgren and Schwartz [25] as *the noticing effect*: 'A characteristic of perceptual learning is the increasing ability to perceive more in a given situation. Experts can notice important subtleties that novices simply do not see…[This] helps explain how people can come to perceive what they previously could not, and how the ability to notice often corresponds to competence in a domain.' (p. 421). This is similar to what Goodwin [26] calls *professional vision*. It is thus 'essential for learning to notice what is important *and* what is not important' [25](p. 426) and that can be done by experiencing appropriate variation [22].

However, this does not address what it is that makes the difference between a novice and an expert and how a novice move from being a novice to becoming an expert or disciplinary insider; here *reflection* play a crucial role for the process of learning. John Dewey characterized reflection as: 'active, persistent and careful consideration of any belief or supposed form of knowledge in the light of the grounds that support it and the further conclusions to which it tends, constitutes reflective thought' [27](p. 9). Building on Dewey's work, Schön [28] modeled reflection in two different ways: *reflection-on-action* and *reflection-in-action*. *Reflection-on-action* occurs when one thinks back on the solving of a problem, whereas *reflection-in-action* occurs when one is aware of, and communicate to others, ones' thoughts on the problem while engaging in solving it. Adding reflection to noticing then characterize changes in one's thinking, which corresponds to the process of learning.

Building on these concepts, I define *disciplinary discernment* as:

> noticing something, reflecting on it, and constructing new meaning from a disciplinary perspective.

It is thus not ordinary discernment referred to here, but discernment of *disciplinary affordances* of representations. *Disciplinary affordances* is defined by Fredlund, Airey and Linder [29] as the 'inherent potential of a representation to provide access to disciplinary knowledge' (p. 658). Fredlund, et al. [29] go on to argue that 'learning then, involves coming to appreciate the disciplinary affordances of representations' (p. 658).

However, when a novice looks at a representation of an astronomical object, which normally includes many disciplinary affordances, the novice will only discern a subset of the total disciplinary affordances, set by the discipline community, of that representation [cf. 30]. Hence, it is important for professors to be aware that students do not discern the same things from a representation as they do. Thanks to the professors disciplinary knowledge and educational background, they have developed professional vision [26] and are 'sensitive to patterns of meaningful information that are not available to novices' [31](p. 33) in that they can both evaluate and criticize representations in an automatic, unconscious manner [32,33]. The novices, on the other hand, often focus on the wrong things. Eriksson, et al. [14] shows that an alarming 43% of the 1st year undergraduates that participated in their study focused on non-disciplinary things in the presented representations, which highlights the educational challenges by the professors.

This realization lead to an investigation of what university students and professors of astronomy discern when they engage with the same disciplinary representations, in this case a simulations of flythrough of our galaxy [14]. The findings from this qualitative study showed large differences in disciplinary discernment in relation to educational level and lead to a disciplinary discernment hierarch – the Anatomy of Disciplinary Discernment (ADD) [14]–'which, from an educational perspective is how disciplinary knowledge can be seen to increase as a function of a growing ability to discern disciplinary crucial aspects from a vast array of potential affordances of a given representation' (p. 179).

In table 1, a representation of the ADD is presented. It is constructed by five levels if increasing discernment, where the first level—Non-disciplinary discernment—involves discernment that has nothing to do with the discipline, similar to everyday discernment. The other levels of discernment refer to increasing disciplinary discernment: the first—Disciplinary Identification—involves recognition and naming of salient disciplinary objects, for example "it is a star!" or "that is a nebula". The second level—Disciplinary Explanation—involves assigning disciplinary meaning to discerned objects, for example "the nebulae is red due to Balmer emission in hydrogen gas". The third level of discernment—Disciplinary Appreciation—involves acknowledging the value of the affordances of the representations, for example "[When I see this] I start to think about all the things I have learned during the course. What a nebula is, how stars are born, supernovae, and other concepts that I have learned. This picture is not entirely like other pictures I have seen on this object". Finally, the forth level—Disciplinary Evaluation—involves analyzing and critiquing, both positive and negative, the representations used for an intended affordance, for example, "There is a lot going on that we can't see at visible wavelengths – this would be a very different clip in infrared light, for example". For more detailed descriptions of the categories, see Eriksson, et al. [14].

Table 1. This table represent the hierarchy of the Anatomy of Disciplinary Discernment (ADD). For details, see Eriksson, et al. [14].

| | **The Anatomy of Disciplinary Discernment** |
|---|---|
| Increasing levels of Discernment | Disciplinary Evaluation |
| | Disciplinary Appreciation |
| | Disciplinary Explanation |
| | Disciplinary Identification |
| | Non-Disciplinary Discernment |

The results from the empirical study clearly showed that disciplinary discernment vary a lot and the most important factor was the educational level by the participants; higher levels of disciplinary discernment correlated with higher educational level within the discipline.

Taking all of this into account, my argument here is that students discern different things in representations compared to the professors, and the ADD can describe the different levels of disciplinary discernment.

I will now address the next fundamental aspect needed for to build *Reading the Sky:* Extrapolating three-dimensionality.

## C. The multidimensionality (MD) hierarch and Extrapolation of thee-dimensionality (E3D)

This section concerns the concept of spatial thinking. Therefore, I start by defining what I mean by spatial thinking in an astronomy education context. Spatial thinking is:

> '*the recognition, consideration, and appreciation of the interconnected processes and characteristics among astronomical objects at all scales, dimensions, and time*' [10](p. 118).

Spatial thinking has increasingly been seen as an important competency in different science disciplines and astronomy is no different in this respect [6]. Indeed, astronomy as a discipline demands excellence in the ability to be able to extrapolate three-dimensionality in one's mind from one- or two-dimensional inputs. The astronomical distances in the Universe offers little, if any, possibility to experience an astronomical object from different directions; consequently, the input to our senses are at best two-dimensional. An additional complication is the distance in itself; most astronomical objects are so distant that they cannot be seen by the naked eye. Consequently, every object in the Universe need a representation. It is from these representations that our understanding of the Universe is built. Usually, these representations are one- or two-dimensional to its nature. Following the definitions by Gilbert, Reiner and Nakhleh [34], by one-dimensional, I here refer to text, symbols and mathematics, whereas two-dimensional representations could be diagram, graphs, images, etc. Three-dimensional representations are, for example, gestures, real physical models or simulations and aminations, where one can move around in virtual reality universe, or use 3D glasses.

It is often taken for granted in astronomy education that students will be able to, in their minds, extrapolate a 3D experience from 2D representations. This is despite the growing body of research indicating that this is often not the case [6,8,35-38]. For example, Parker and Heywood [37] found that students had great difficulty in moving from 2D representations of the solar system to 3D representations. They concluded that 'there is a generic problem of spatial awareness in relating to position in space of the observer and the observed objects.' [37](p. 515). However, by using simulations and animations, where students can manipulate objects and positions, others have found that students learn more effectively about astronomical concepts by letting them becoming 'living phenomena that are *actualized* and not simply *realized*' [39](p. 751). This highlights the challenge the students face when entering the astronomy discipline; to learn to understand the Universe, they need to learn to think spatially, or develop their ability to extrapolate three-dimensionality from one- or two-dimensional inputs. However, it is not clear from the astronomy education research literature what the required spatial thinking entails and is used for, but experienced astronomy professors repeatedly observe successful students possessing high levels of spatial thinking [7]. The lack of literature in the field is concerning but there is a growing interest from AER community to further investigate the connection between success in learning astronomy and extrapolating three-dimensionality [40-43].

Extrapolating three-dimensionality is thus identified as important in astronomy education, but to what extent do students discern dimensionality aspects in representations? In an empirical qualitative study on discernment of dimensionality from simulations Eriksson, et al. [6] were able to show that the ability to extrapolate three-dimensionality vary by the participating students and professors. The discernment of dimensionality could be categorized into a multidimensionality (MD) hierarchy of six categories of disciplinary discernment, which could be clustered as a function of 1-3 dimensionality thinking, see table 2 and Eriksson, et al. [6].

Table 2. The multidimensionality (MD) hierarchy. This hierarchy is built by six categories of discernment of dimensionality thinking. For details, see Eriksson, et al. [6].

| | Categories of discernment ordered hierarchically | Clustering of categories as function of 1-3 dimensionality |
|---|---|---|
| Categories of disciplinary discernment | Advanced three-dimensionality | Three-dimensionality |
| | Growth of three-dimensionality | |
| | Emergence of three-dimensionality | |
| | Relative size awareness | Two-dimensionality |
| | Distance contemplation | One-dimensionality |
| Baseline category | Motion identification | |

The one-dimensionality disciplinary discernment cluster involves discernment of motion, i.e. the experience of moving forward in terms of direction, speed, acceleration, or rotation. Also, discernment of distance travelled or distance between objects and direction falls under this cluster. The two-dimensionality cluster consists of only discernment of relative size between astronomical objects. For the three-dimensionality cluster, three different type of disciplinary discernment related to dimensionality awareness were identified. These are the 'emergence of three-dimensionality awareness category—discerning structures and details within astronomical objects—includes a level of awareness of depth, i.e., a third dimension. The growth of three-dimensionality awareness category includes an awareness of how different astronomical objects change appearance due to motion parallax or a change of position perspective [. Finally] the advanced three-dimensionality awareness category involves a more complete discernment of three-dimensionality in relation to the structure of the universe, and therefore is the most sophisticated category level.' [6](p. 431). These three discernment categories all incorporate discernment of both distance, width and depth, with reflects a 3D awareness of the Universe. For a more detailed description of the categories, see Eriksson, et al. [6].

For the arguments presented in this paper, it is discernment related to the three-dimensionality cluster that is most relevant, as it demands the competency to extrapolate in one's mind a 3D awareness of the Universe.

Finally, I turn to the concept of *Reading*.

## D. Reading

Metaphorically, to *read* something has many meanings and applications, besides the obvious of reading a written text. For example, cultural geographers commonly talk about *reading the landscape* [44-47], ecology educators talk about *reading nature* [48], and others uses the term *Reading science* to discuss meaning-making and communication from a social semiotic and semantic approach [4]. From a more general perspective, Card, Mackinlay and Shneiderman [49] and Kress and van Leeuwen [5] theorizes around reading visualizations or images. For the purpose presented here, I draw on the metaphors of *reading the landscape* and *reading nature*.

For cultural geographers, *reading the landscape* concerns the ability to "see" the landscape in the kind of disciplinary way that facilitates the generation of insightful understanding. Hence, the usage of the term calls for a disciplinary understanding of the "language" of landscapes [47]. This means that *reading the landscape* metaphorically symbolizes the interpretation of a given piece of landscape from observations as if one was reading the "text" of cultural geography "language". I interpret such use of *reading the landscape* as an example that vividly captures how disciplinary-specific representations get used to share perceptions, knowledge and meaning-making. In cultural geography the landscape is seen as 'being *always already* a representation' [47](p. 68), which, by virtue, is visually three-dimensional in nature. This framing is useful for making the case for the idea of *Reading the Sky*. Consider the resemblances between the notions of *reading the landscape* and *Reading the Sky*: both the landscape and the sky need to be observed and to make sense of those observations they need to be "read" using an appropriate disciplinary "language" [cf. 4,50]. Learning such "language" is essentially what the educational endeavour is about in any discipline [see, for example, 51,52]. For example, in cultural geography Wylie [47] says: '*reading* refers largely to knowledgeable field observations, and where the landscape is a book in the broadest sense' (p.71, emphasis added). As such, landscapes are representations to cultural geographers that are to be interpreted rather than just described. Since the ability to read a landscape must vary, the interpretations of what is observed must vary: 'There is no single, "right" way to read a landscape' [44](p. 603). However, cultural geography educational literature offers little guidance on how fluency in reading the landscape can be educationally achieved.

The educational framing for *reading nature* by Magntorn [48] in ecology education is, however, more developed and I thus find this framework to be a good starting point to establish the framing of *Reading the Sky*. Magntorn describes how *reading nature* involves two important elements: first, *discernment*, which he defines as being 'able to see things in nature and to discern the differences and similarities between objects in nature' [48] (p. 17) and second, *discussion*, which for him is effective communication using disciplinary-specific multimodal representations. These two aspects are interconnected with "outdoor experiences" and "theoretical knowledge" regarding, for example, organisms, processes, and abiotic factors, i.e., becoming fluent in the disciplinary discourse of ecology [cf. 3]. Furthermore, Magntorn frames his findings in terms of what he calls "competence", which he characterizes in terms of content knowledge and its associated attained proficiency. In so doing, Magntorn proposes a revised Structure of Observed Learning Outcomes (SOLO) taxonomy [53,54]. This revision describes different levels of sophistication concerning *reading nature* from an ecology education perspective. The levels are used to classify students' and teachers' ability to read nature, and to discuss critical aspects for learning to read nature from a phenomenographic point of view [22,55,56].

## III. BUILDING READING THE SKY

### A. Defining *the Sky*

At this point in my discussion I need to further explain what I mean when I refer to *the Sky* [10]:

> *The Sky* is the whole Universe at all levels of detail, including all forms of disciplinary-specific representations, and other semiotic resources, describing the Universe, at all scales, its properties, but also the processes involved in their interaction with the surrounding, at local scale and large scale, and time.

This highlights the size of the challenge at hand; to be able to competently get to "see" the whole Universe, its parts, and how they interact. Obviously, from an educational point of view, the competency to read *the Sky* must be seen against the educational aims of a given educational context. I come back to this in a later section.

### B. Defining *Reading the Sky*

I now propose a definition of *Reading the Sky* that brings together the metaphor "Reading" with the extrapolation of three-dimensionality and disciplinary discernment [10]:

> *Reading the Sky* is the competency to discern disciplinary affordances of *the Sky* in order to acquire a holistic, three-dimensional, understanding of the Universe at all levels of scale, dimensions and detail.

In the definition of *Reading the Sky*, I draw on the concepts of *reading the landscape* [47] and *reading nature* [48], capturing the varying ability to discern and interpret how disciplinary-specific semiotic resources get used to share perceptions, knowledge and meaning-making within a discipline. Both of these involve observations and measurements, which have great importance for all of astronomy, but also how these are perceived. As such, *Reading the Sky* is grounded in disciplinarity and bridges the gap between the discipline of astronomy and the meaning-making that gets constructed from observations and measurements by astronomy learners. *Reading the Sky* thus concerns disciplinary discernment of any representation belonging to the astronomy discourse. However, the discerned disciplinary affordances for a particular representation [29] will only constitute a subset of those set by the discipline [cf. 30]. Hence, there is a potential risk of students missing educationally relevant aspects. Part of the reason for the limitations to what a student can discern from disciplinary representations comes from cognitive load theory [see, for example, 57]. Cognitive load theory portrays the amount of information that can be perceived through vision as being not only limited *per se*, but also limited by information perceived by our other senses. This becomes a particularly important consideration when choosing to use simulations and animations, that attempt to realistically represent aspects of nature, as a teaching tool. Here, there is a potential risk of students missing educationally relevant aspects because of cognitive overload [57,58] or, by only focussing on the most visually compelling attributes, which might not be relevant for the task at hand [22,59-63]. Astronomy students will need to first discern and then unpack the disciplinary affordances of the representations, which have been found to be both difficult and problematic for many students [6,14,64,65].

*Reading the Sky* observations thus include what one can discern using telescopes, by looking at spectra from stellar objects, from images of the sky and astronomical objects or, from any kind discipline-based representations. All these use the naked eye as a detector. Discernment is constructed from these observations through a meaning-making process that calls for a "fluency" [3] in disciplinary discourse, which is linked to spatial thinking [6]. Becoming part of the discourse of astronomy thus involves being able to "fluently" "read" *the Sky* by interpreting, understanding and using the different semiotic resources, and in particular the representations, that astronomers use to communicate disciplinary knowledge as part of developing a discursive identity [24,66,67]. Using this disciplinary discourse perspective, *Reading the Sky* calls for the two abilities, "disciplinary discernment" and "extrapolating three-dimensionality" to be linked to "observations and experiences" and "disciplinary knowledge" in order to be able to "see" through vision, and "interpret" through the affordance of disciplinary-specific representations, the Universe. This is illustrated in Figure 1.

## IV. READING THE SKY – A COMPETENCY

In this paper, it is argued that becoming part of the discourse of astronomy involves being able to fluently *Read the Sky* by interpreting, understanding and using the many different representations that astronomers use to communicate disciplinary knowledge. From the ADD and MD hierarchies [6,14], I propose that *Reading the Sky* can be seen as a competency highly important for learning astronomy. However, as opposed to the SOLO, or Bloom's taxonomies [68-71], the framework of *Reading the Sky* is grounded in disciplinarity and builds on appreciating how disciplinary knowledge relates to actual astronomical observations of *the Sky* as a function of discerning the disciplinary affordances of representations used in the disciplinary discourse of astronomy. For people not in the field of astronomy, or not having a strong personal interest in astronomy, the general level of astronomical knowledge can be expected to be low. However, for university students taking courses in astronomy, *Reading the Sky* becomes a competency that needs to be striven for; the framing of *Reading the Sky* provides a research-informed link between observations and the meaning-making that gets constructed from those observations. Therefore, *Reading the Sky* becomes important when addressing learning astronomy, and literature reviews reveal that this has not been addressed or quantified earlier. As indicated earlier, I am now doing this from a disciplinary discernment perspective. Through my long experiences of teaching physics and astronomy and carrying out research in both astronomy and astronomy education, I have become fascinated by what students/observers actually believe that they "see" – observe, discern, or *read* – when watching the real sky, visualizations of the Universe, or any other astronomy representations. The disciplinary ability to *read the Sky* is thus very complex in nature and so I propose that *Reading the Sky* should be quantified as a competency and used as an effective astronomy education tool.

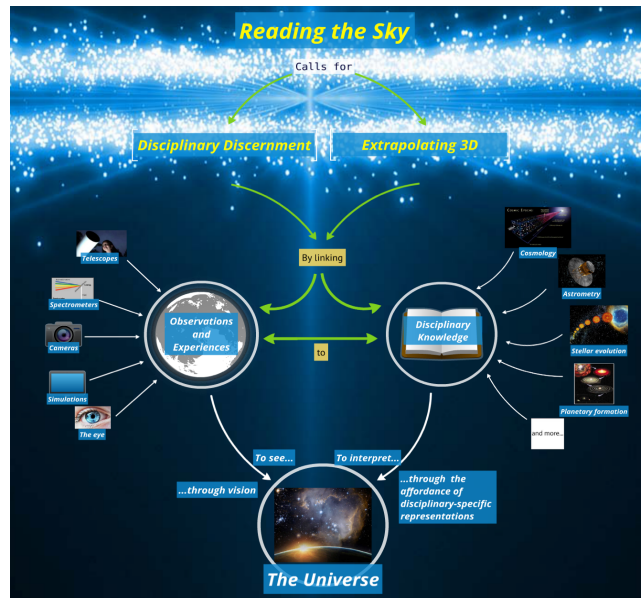

Figure 1. A graphical representation of how *Reading the Sky* is related to disciplinary discernment and extrapolating three-dimensionality, by linking observations and experiences to disciplinary knowledge.

Complementary to disciplinary knowledge, competency in *disciplinary discernmen*t and *extrapolating three-dimensionality*, i.e. handling disciplinary knowledge in appropriate ways using disciplinary-specific representations, can be identified as being important for *Reading the Sky*. Of course, these competencies can only be theoretically separated; in practice they are intertwined with disciplinary knowledge, theory, and practice [15,32]. As such, *Reading the Sky* opens up a new way to expand learning astronomy through disciplinary discernment by an inclusion of spatial thinking, i.e. *extrapolating three-dimensionality* from one or two-dimensional input. The discernment of relevant structural components of the Universe and how they interact through different processes, involves looking at, reflecting on, and constructing meaning, in relation to the whole of the multidimensional Universe. This is done through observations and measurements, which have great importance for all of astronomy. Therefore, *Reading the Sky* works as a bridge between the discipline of astronomy and astronomy education.

At this point, it is important to recognize that it is people making and interpreting these observations; hence they cannot be absolutely objective [72,73] and rely heavily on what a person experience [74]. In all scientific activity, this phenomenon is well-known; for example, the natural scientist Alexander von Humboldt [75] wrote about in his *Aspects of Nature* as early as 1849. The dark sky, distant light, colours, odours and fragrance from afar, the silhouette of the horizon, sound, etc., collectively contributes to what gets "read" by an observer. The challenge lies in only observing the relevant features [76] and the challenge behind that is knowing what these relevant features are and/or how to recognise what these are. This becomes increasingly more important in situations where one needs to rely heavily on one's eyes to make observations [23].

From an epistemological point of view, all knowledge or well-grounded belief rests on experience; experience that is gained through direct sense-perception, in this case, vision. Furthermore, since none of us have identical prior knowledge, different people discern identical things differently [26]. Again this is not a new idea, it is well known that Socrates used this to argue that our senses cannot access reality in any direct way, or, as Shapere [76] express it: 'sense-perception is notoriously untrustworth' (p. 508). At the same time, an image of, for example, a nebula taken by a CCD-camera through a telescope, see figure 1, is also only a representation of the "real" object. This kind of representation gets built on chains of representations that have been coordinated by people, often over a long time period [64]. These kinds of representations have intended meanings, created by someone with a particular purpose, and made up by an enormous number of disciplinary affordances [29]. Typically, only a subset of the disciplinary affordances will be discovered and/or discerned by students (and even their teaching professors) [65], and this discernment can be quantified and described by the ADD and MD hierarchies.

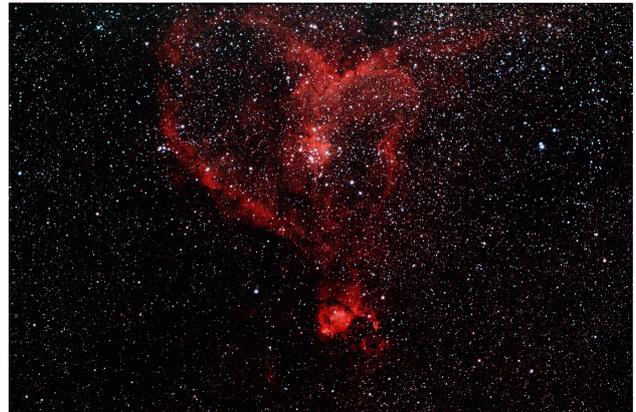

Figure 1. A CCD-image of the "Heart Nebula", IC 1805, taken by a telescope. It is important to be aware that this is only a representation of the real object. Image credits: Jonas Carlsson

In the next section, I propose a theoretical model for how to use *Reading the Sky* by characterizing what is needed to link *extrapolating three-dimensionality* and *disciplinary discernment* to *disciplinary knowledge* as part of informing the optimization of the teaching and learning astronomy.

## V. THE SPIRAL OF TEACHING AND LEARNING IN ASTRONOMY

I will now turn to show how *Reading the Sky* can be used to inform teaching and learning astronomy by framing it around two powerful educational ideas. First, the 'Spiral Curriculum' [77] and second, 'Visible Learning' [78,79].

The Spiral Curriculum idea, proposed by Bruner, involves information being structured so that complex ideas can be taught at a simplified level first, and then re-visited at more complex levels later on, gradually increasing the disciplinary representation affordances, hence the spiral analogy. However, Bruner does not explicitly address how this could be done. Following Eriksson, et al. [14], it would involve using certain disciplinary representations in a teaching sequence and first start at identifying different salient aspects in these representations. Then, by discussing them and adding information, students would learn and, when revisiting the representations again, be able to discern more disciplinary affordances, which potentially could lead to enhanced learning or open pathways to new knowledge. Ideally, this would have the students to cross over category boundaries in the ADD, and, by explicitly addressing the dimensionality aspects, also have them cross category boundaries in the MD hierarchy. Using this as a starting point for a model for disciplinary learning, similar to growing into the discipline, I propose a three-dimensional spiral space of learning – *the Spiral of Teaching and Learning*, see figure 2. According to this model, student can learn any material so long as it is taught and organized appropriately. From a constructivist point of view, learning is a process that involves con-structing knowledge oneself and/or in groups by organizing and categorizing information. By providing appropriate educational experience that follows sequencing based on both the ADD and MD hierarchy, students may learn more effectively from *Reading the Sky* as they discover the disciplinary ways of how the discourse is organized and communicated.

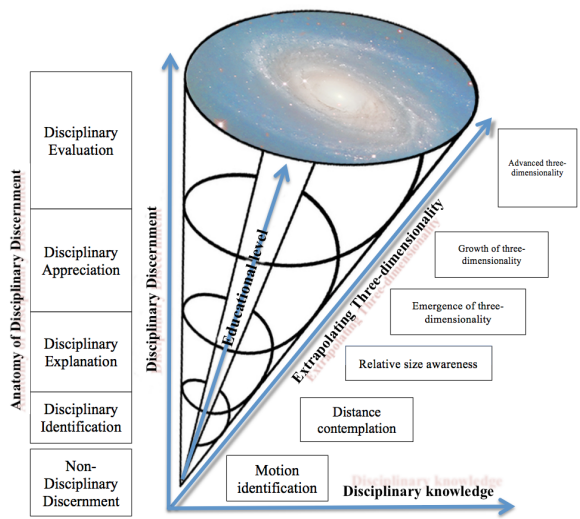

Figure 2. This idealized representation illustrates Reading the Sky as constituted by three abilities: Disciplinary discernment, Extrapolating three-dimensionality, and Disciplinary knowledge. I refer to this three-dimensional space as the Spiral of Teaching and Learning'. Since these abilities are intertwined with each other, there are numerous possible learning trajectories for the teacher to consider.

However, this is not enough. Students need help in this process of disciplinary discernment and deduction and guidance by a teacher is required [80]. Hattie's [78,79] ideas of visible learning highlights the importance of the teacher in this process: 'It is teachers seeing learning through the eyes of students [and,] the greatest effects on student learning occurs when teachers become learners of their own teaching, and when students become their own teachers' [79](p. 14). This is important in that the teacher must first find out where the students are in ADD and MD and with that knowledge as a starting point plan the teaching by helping the students to discern relevant disciplinary affordances of representations [cf. 81,82]. One must remember that teachers, or experts in general, often have forgotten how difficult it was to discern relevant disciplinary affordances of representations before they became experts [26,31-33,83]. This framework thus provides the clear connection between the abilities needed for competently both teach and learn astronomy. Learning astronomy involves becoming competent in all three abilities, and research has shown that novice learners have very limited abilities when it comes to both disciplinary discernment and extrapolation of three-dimensionality [6,7,14,84]. Consequently, teaching astronomy must involve much more than disciplinary knowledge; the astronomy teacher needs to take into account both disciplinary discernment and the ability to extrapolate three-dimensionality from one- or two-dimensional representations, which takes lots of practice to competently master [6], by providing learning situations and exercises that particularly address all these issues, using appropriate variation [22,56]. To do that, teachers must first probe for where the learners are in the discernment hierarchies before starting any teaching sequence [cf. 81]. As a result, the role of the teacher becomes very important in generating the scaffolding needed to help students *cross over category boundaries* in the hierarchies in *the Spiral of Teaching and Learning* in achieving *Reading the Sky* - competency.

## VI. DISCUSSION AND IMPLICATIONS

From an educational point of view, *Reading the Sky* competency must be seen against the educational aims of a given educational context. For many students, learning astronomy may begin with local observations of day and night skies. During the day, the Sun and the Moon could be visible. At night the Moon, planets, stars, nebulae, and galaxies could be visible. Observing the night sky in a planetarium or through simulations on computers, tablets, or smart-phones, could educationally enhance such observations. My point is that the aim of astronomy courses should be set against being able to *read the Sky* sufficiently well in terms of the aims and objectives of the course. This can be seen as achieving a designated level of 'literacy' in the relevant parts of the disciplinary discourse of astronomy applicable to a given course [cf. 85,86]. In this way, achieving competency in *Reading the Sky* involves traversing my adaption of Bruner's model of the spiral curriculum [77], which in turn carries similarities to the historical development of today's accepted astronomical view [87-92]. As such, I argue that *Reading the Sky* competency is vital for efforts aimed at optimising and improving astronomy education.

Developing *Reading the Sky* competency by astronomy students calls for careful consideration by an astronomy professor. From the results of Eriksson, et al. [6] it is argued that simulations that promote the extrapolation of three-dimensionality have the distinct possibility to help students build the curriculum-intended level of understanding of the three-dimensional Universe. However, the ability to extrapolate three-dimensionality depends strongly on the role that the representations play in simulations to provide the needed motion parallax; one of the most important findings from that study is that motion parallax cues the ability to extrapolate three-dimensionality. This, in itself, depends on disciplinary discernment and here astronomy professors need to generate the scaffolding needed to help the students discern the relevant things and thus *cross over category boundaries* in the ADD. So, the tailoring of teaching sequences to provide students with opportunities to develop their abilities concerning spatial thinking in terms of extrapolating three-dimensionality through disciplinary discernment is educationally critical. The role of the professor needs to be much more than a "guide" that takes the students through the chosen simulation or other semiotic resources. What is needed is a teaching mind-set that takes the students 'on an excursion into the target discourse arena, gradually shifting the frame of reference until it corresponds well enough to allow sense to be made within the specialist discourse' [93](p. 263). This requires a teaching approach that signals the mapping between disciplinary-specific representations without over-burdening the students by making the task too complex [62]. Hence, such professors, by making boundary crossing in the ADD and MD hierarchies possible (see table 1 and 2), will be tailoring their teaching to provide the necessary scaffolding to help students to learn to extrapolate three-dimensionality as a function of developing spatial thinking. The theoretical model of the *Spiral of Teaching and Learning* in astronomy provide and describe such framework by visually bring to the fore how disciplinary discernment, spatial thinking and, disciplinary knowledge are necessary abilities for the creation of an optimal spiral of teaching and learning. Further, set against the idea of achieving competency in *Reading the Sky*, I would argue that the three intertwining competencies in my proposed *Spiral of Teaching and Learning*, provide the essential grounding for the generation of the scaffolding needed to optimize the teaching and learning of astronomy. From my experience, traditional astronomy teaching focuses on achieving learning disciplinary knowledge and practices. As such, it assumes that the kind of educational challenge that my research has bought to the fore gets taken care of as a natural part of the associated learning. Considering Eriksson's, et al. [14] finding, that almost half of first year students were making non-disciplinary discernment, clearly shows this is not a valid educational assumption. Therefore, teacher awareness is needed about the crossing of category boundaries in all 'dimensions' in the *Spiral of Teaching and Learning* as a significant step to establish learning that increases in sophistication as educational experience progresses. This is not straightforward, since what is obvious for professors may not even be discernable for students in or entering into the discipline of astronomy [see, for example, 6,14,31,93-95]. Thus, the role of the teacher is critical here and I suggest that Hattie's [78,79] idea of visible learning offers a pragmatic way to think about the crossing of category boundaries in all "dimensions" in the *Spiral of Teaching and Learning*.

The *Spiral of Teaching and Learning* have been extensively used in my teaching of astronomy over the years and is found to offer a working model for how to effectively teach astronomy. Also, students report enhanced understanding and an awareness of what semiotic resources (representations, activities and tools) mean for their learning and the discipline. To exemplify how I work with this in an astronomy education context, I provide two examples below. Both examples use simulations as semiotic resources, which includes a number of disciplinary-specific representations and numerous affordances. The first example is taken from a course in planetary astronomy and involves gravity and orbits. The second example is from an introductory course discussing phases of the Moon. Both examples start with the same first phase:

1) Start by constructing or finding a suitable semiotic resource that you think clearly and realistically represents aspects of the disciplinary knowledge you plan to teach. In these cases one can use, for example, a PhET simulations (https://phet.colorado.edu/sims/html/gravity-and-orbits/latest/gravity-and-orbits_en.html) or a YouTube video (https://youtu.be/jgoIP90apEs).
2) Evaluate the chosen semiotic resource for its disciplinary affordances.
3) One must now 'become a student' and try to envision what your students will discern from the simulation.
4) Present the semiotic resource for the students and ask for their discernment using, for example, an 'exit ticket'[*] or as a simple web-based questionnaire, after a prior lecture.
5) Analyse the answers using the ADD and MD hierarchies to find the students levels.
6) Plan you teaching using this information.

The second phase depends on your chosen content. In both examples presented here, the chosen semiotic resource will be central for the lecture.

For gravity and planetary orbits:
7) Discus the experience of orbits and what the students discern/identify in the semiotic resource. Point to what you think is most important and relevant for the chosen curriculum.
8) Help the students to discern how planets orbit at different distances from its star. Let them explore how orbits can be expressed and explained, using physics, to construct a model, i.e. they start to become their own teachers. This expanded disciplinary knowledge may enable the students to discover more disciplinary affordances by the simulation when revisiting it again and build their competency in *Reading the Sky*.
9) Using this expanded disciplinary discernment, apply this new expanded disciplinary knowledge by adding more planets and moons or using other simulations/resources. The students may be able to discern and appreciate more disciplinary affordances by the semiotic resources and start to construct advanced disciplinary knowledge. i.e. increasing their *Reading the Sky* competencies.

For phases of the Moon:

---

[*] For examples on the use of exit tickets, see Brown University: https://www.brown.edu/about/administration/sheridan-center/teaching-learning/effective-classroom-practices/entrance-exit-tickets

7) Discus the experience of the Moon and its phases and what the students discern/identify in the semiotic resource. Point to what you think is most important and relevant for the chosen curriculum, including spatial thinking.
8) Using the semiotic resource, help the students to discern the different phases and how they are presented in the resource. Have them explore how this can be used to understand how the phases of the Moon arises and build a conceptual model for this phenomenon. Now, they can explore the semiotic resource again for more disciplinary affordances.
9) Let the students explore other semiotic resources showing the same phenomenon in 3D and help them discern new affordances, thereby creating new or enhanced knowledge, meaning and *Reading the Sky* competency. Finally, the students may appreciate the phenomenon presented by the semiotic resources and in one's mind being able to extrapolate a three-dimensional understanding of the phases of the Moon.

Of course, many more examples could be given, using other semiotic resources, such as representations, real observations of the sky, or physical models. I chose to present cases using simulations, since these have been found useful for teaching and learning astronomy by often offering educational relevant affordances and, most importantly, motion parallax for building expanded awareness for extrapolation of three-dimensionality [see, for examples and overviews, 25,96,97]. For the example with the phases of the Moon, these exercises should be supported by real naked-eye observations by the Moon, to further experience what it looks like on the sky.

## VII. CONCLUSIONS

The aim of this paper was to introduce and discuss two new concepts for teaching and learning astronomy: *Reading the Sky* and *the Spiral of Teaching and Learning* in astronomy. In this paper, I have argued that becoming part of the discourse of astronomy involves being able to fluently *read the Sky* by discerning, interpreting, understanding and using the many different semiotic resources that astronomers use to communicate disciplinary knowledge. Using a disciplinary discourse perspective [3], *Reading the Sky* calls for the two abilities, disciplinary discernment and extrapolating three-dimensionality, to be linked to observations and experiences and disciplinary knowledge in order to be able to see through vision, and interpret through the affordance of disciplinary-specific representations, the Universe. By using the ADD and MD hierarchies [6,14], I propose that *Reading the Sky* can be seen as a competency highly important for learning astronomy. As such, *Reading the Sky* competency is vital for efforts aimed at optimising and improving astronomy education.

By framing *Reading the Sky* around two powerful educational ideas, the Spiral Curriculum [77] and Visible Learning [78,79], I introduce the *Spiral of Teaching and Learning,* a theoretical model for how to inform teaching and learning in astronomy. This model, where disciplinary discernment and spatial thinking are equally important as disciplinary knowledge, have been found effective for teaching and learning astronomy at university level. It can be assumed that the model also applies to other science subjects, particularly where one cannot easily see different phenomena, such as atomic interaction in chemistry, or in atomic and particle physics where the situation is similar to astronomy in that representations are needed to visualize various phenomena. The most important argument presented in this paper is that the proposed theoretical framework should be seen by astronomy professors to offer a new way of planning teaching and learning spirals to enable students to become part of the disciplinary discourse of astronomy: students need to learn to develop *Reading the Sky* competency, or else they will only see and not discern. The distinction is clear.

## ACKNOWLEDGEMENT

Many thanks to Prof. Ann-Marie Pendrill and Prof. Andreas Redfors for comments and suggestions in improving this paper. Also, thanks to Prof. Ed Prather for good discussions and ideas around the concept of Reading the Sky. Finally, thanks to the anonymous reviewers for their constructive suggestions.